\documentclass[prb,twocolumn,superscriptaddress,showpacs]{revtex4}
\usepackage{graphicx}
\newcommand{\beq}{\begin{equation}}
\newcommand{\eeq}{\end{equation}} 
\newcommand{\beqa}{\begin{eqnarray}}
\newcommand{\eeqa}{\end{eqnarray}}
\newcommand{\ba}{\begin{array}}
\newcommand{\ea}{\end{array}}
\begin{document}
\bibliographystyle{prbrev}

\title{Vortex arrays in nanoscopic superfluid helium droplets}

\author{Francesco Ancilotto}
\affiliation{Dipartimento di Fisica e Astronomia ``Galileo Galilei''
and CNISM, Universit\`a di Padova, via Marzolo 8, 35122 Padova, Italy}
\affiliation{ CNR-IOM Democritos, via Bonomea, 265 - 34136 Trieste, Italy }

\author{Mart\'{\i} Pi}
\affiliation{Departament ECM, Facultat de F\'{\i}sica,
and IN$^2$UB,
Universitat de Barcelona. Diagonal 645,
08028 Barcelona, Spain}

\author{Manuel Barranco}
\affiliation{Departament ECM, Facultat de F\'{\i}sica,
and IN$^2$UB,
Universitat de Barcelona. Diagonal 645,
08028 Barcelona, Spain}

\begin{abstract}
We have studied the appearance of vortex arrays
in a rotating $^4$He nanodroplet at zero
temperature within density functional theory. 
Our results
are compared with those for classical rotating fluid drops used to
analyze the shape and vorticity in recent  experiments 
[L.F. Gomez et al., Science {\bf 345}, 906 (2014)],
where vortices have been directly seen in superfluid droplets 
for the first time.
In agreement with the experiments, we have found 
that the shape of the droplet changes from
pseudo-spheroid,
oblate-like for a small number of
vortices to a peculiar
``wheel-like" shape, delimited
by nearly flat upper and lower surfaces, 
when the number of vortices is large. 
Also in agreement with the experiments, we 
have found that the droplet remains
stable well above the stability limit predicted 
by classical theories.

\pacs{67.25.D-, 67.25.dk, 67.25.dr}
\end{abstract}
\date{\today}
\maketitle

Helium-4 droplets
created by expanding a cold helium gas\cite{Toe04} or fragmentation of a
criogenic liquid attain a limiting
temperature below 0.4 K,\cite{Har95} and constitute the only
self-bound superfluid systems. Superfluidity in
helium droplets was established through the
dissipationless rotation of
an OCS molecule inside them, as indicated by the appearance
of a clean ro-vibrational spectrum.\cite{Gre98}
More recently, the indirect evidence 
of quantum vortices\cite{Gom12,Lat14,Tha14} and the
frictionless displacement of swift impurities in helium 
droplets \cite{Bra13} point towards a superfluid character of helium nanodroplets.

Superfluid $^4$He droplets cannot be set into rotation as ordinary
droplets or rigid bodies. If a rotating helium
droplet in the normal phase above the superfluid 
transition temperature $T_{\lambda} =
2.17$ K is cooled down reaching the superfluid phase,
it reacts by storing its angular momentum 
either into quantized vortices
or into travelling capillary
waves.\cite{Lea14}
Conversely, a critical angular velocity $\omega_c$ has to be supplied to
the superfluid droplet for the nucleation of
vortices with quantized velocity circulation in units of $h/M$, where
$h$ is the Planck constant and $M$ is the mass of a $^4$He atom.
Single vortices in helium droplets
have been addressed theoretically 
by methods of different complexity, see e.g.
Refs. \onlinecite{Bau95,Dal00,Leh03,Anc03,Sol07}. 

When the angular velocity is increased above $\omega_c$, larger
amounts of angular momentum may be stored
into the superfluid by increasing the number of nucleated vortices.
These vortices arrange themselves into ordered structures (lattices)
whose existence in bulk superfluid $^4$He was established long 
ago.\cite{Vin61,Wil74}
We refer the reader to Refs. \onlinecite{Don91,Pit03,Fet09} for a
general presentation of the subject.

Very recently, superfluid He nanoscopic droplets in fast rotation
have been studied by coherent X-ray scattering.\cite{Gom14}
The existence of vortex lattices 
inside the droplets was established by the appearance
of Bragg patterns from Xe clusters trapped in the vortex cores 
in droplets made of $N= 10^8 - 10^{11}$ atoms
(corresponding to radii from 100 to 1000 nm) 
produced by the fragmentation of liquid
helium expanding into vacuum.
The shapes of the droplets were consistent with those of axially
symmetric oblate 
pseudo-spheroids
with large aspect
ratio ($AR$), defined as the ratio of the long half-axis length $b$ to the
short half-axis length $a$ along the rotational axis. 
While normal liquid drops change their
shape as rotation becomes faster\cite{Cha65,Bro80,Hil08} to resemble
a ``peanut'' (multi-lobe shape) or a ``blood cell'', no evidence of such
shape shifting has been seen in the helium nanodroplets.\cite{Gom14}
As shown in the following, this is fully confirmed by our calculations.    

The presence of dopants was instrumental
for detecting the vortex cores, although their number was sensibly smaller
than the number of helium atoms ($N_{\rm Xe} \sim 10^{-3} N_{\rm He}$)
and their presence is not expected to introduce large deformations in the droplet
in spite that they locally distort the superfluid around them, see e.g. 
Refs. \onlinecite{Dal00,Anc14}. Possible effects on the
distribution of vortex cores inside the droplet might come
from the additional rotational energy associated to the Xe mass,
especially at the periphery of the droplet. 
Although such effects seem to have been observed
occasionally in the experimental images of Ref. \onlinecite{Gom14},
we will not consider them here. In the case of a rotating nanocylinder, these
distortions were found to be negligible.\cite{Anc14}

Once experimentally established the presence of a vortex lattice in a
droplet of aspect ratio $b/a$, the number $N_v$ of vortices in the
lattice could be determined approximately from the vortex areal density
 (Feynman's formula\cite{Fey55})
 
\begin{equation}
n_v\equiv N_v/{\cal S} =2M \omega /h
\label{eq1}
\end{equation}
where $\omega$ is the rotational angular velocity and ${\cal S} =\pi\, b^2$ is
the equatorial cross section
of the droplet.
Since $\omega$ cannot be directly
determined in the experiment,\cite{Gom14} the analysis relies on the classical
relationship between the $AR$ --experimentally accessible
through the diffraction contour maps-- and
the angular velocity, whose connection with the
parameters $a$ and $b$  is given
by the classical theories of rotating liquid drops.\cite{Cha65,Bro80} 
In this way the experiments estimated that the number of 
vortices in a single droplet could be as large 
as $N_v = 160$.
When the vortex density is
particularly large the experimental images showed also the occurrence of 
``wheel-shaped'' droplets\cite{Gom14}
which have no classical counterpart.

The distinct features of superfluid helium, namely its irrotational flow and 
the possible appearance of quantized vortices, are of course not included in 
the classical rotating droplet model.\cite{Cha65,Bro80}
The existence of a large vortex lattice might influence the appearance of the
rotating droplet, and the irrotational moment of inertia is known 
to be very different from that of the rigid body.\cite{Boh75}
These facts call for theoretically addressing rotating 
helium droplets with accurate methods
which have proven to provide
reliable results for superfluid $^4$He in confined geometries.

We present here a Density
Functional Theory (DFT) study at zero temperature of pure
superfluid helium droplets hosting an increasing 
number of vortices. 
To our knowledge, this is the first realistic study
of multi-vortex configurations in $^4$He nanodroplets.
A previous attempt to study multi-vortex configurations in 
superfluid droplets is described in Ref. \onlinecite{Nam96},
where a simplified model assuming linear vortices and
a rigid spherical droplet was used.

We have
recently analyzed a simpler model system, namely a rotating superfluid $^4$He
nanocylinder hosting arrays of linear vortex lines,\cite{Anc14} 
that constitutes the starting point of the present study.  Within our approach, 
a self-bound superfluid $^4$He droplet is described by a
complex effective wave function $\Psi( \mathbf{r},t)$
related to its atomic
density as $\rho (\mathbf{r},t)= |\Psi( \mathbf{r},t)|^2$.
In the fixed-droplet frame of reference (corotating frame) we seek for
stationary solutions 
$\Psi (\mathbf{r},t) = e^{-\imath \mu t / \hbar} \Phi (\mathbf{r})$,
where  the chemical potential $\mu$ and the time-independent 
effective helium wave function $\Phi$
are obtained by solving the time-independent equation

\begin{equation}
[\hat {H}-\omega \hat{L}_z] \,  \Phi  (\mathbf{r})  =  \,\mu \,
\Phi (\mathbf{r}) \;,
\label{eq2}
\end{equation}
where $\hat{H}$ is the DFT Hamiltonian,\cite{Anc05}  $\hat{L}_z$ is
the angular momentum operator around the $z$-axis, and $\omega$ is the
angular velocity of the corotating frame.

\begin{figure}[t]
\includegraphics[width=1\linewidth,clip=true]{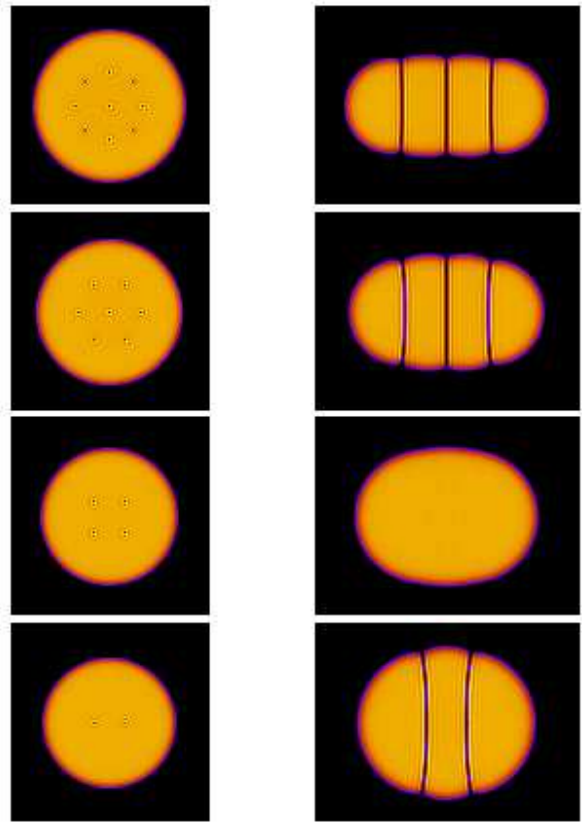}
\caption{\label{fig1} 
Helium droplet configurations hosting (from bottom to top) $N_v=  2, 4, 7,$ and 9 
vortices. The left column shows the density in the $z=0$ 
symmetry plane (top view), while the right column shows side views ($x=0$ plane). 
}
\end{figure}

To determine  $\Phi(\mathbf{r})$ describing a configuration where
$N_v$ vortex lines are present we follow the ``imprinting''  strategy,
 i.e. we start the imaginary-time evolution of Eq. (\ref{eq2})
leading to the minimum energy configuration with a helium wave
function\cite{Anc14}

\begin{equation}
\Phi_0(\mathbf{r})=\sqrt{\rho_0(\mathbf{r})}\, \sum _{j=1}^{N_v} \left[ {(x-x_j)+i
(y-y_j) \over \sqrt{(x-x_j)^2+(y-y_j)^2}}  \right] 
\label{eq3}
\end{equation}
where  $\rho_0(\mathbf{r})$ is the density of the vortex-free droplet
and $(x_j, y_j)$ is the initial position of the $j$-vortex linear 
core with 
respect to the $z$-axis of the droplet.
During the functional minimization
the vortex positions will change
to provide, at convergence, the lowest energy 
vortex configuration. 
It is worth stressing that we work in Cartesian coordinates and
that no symmetry is imposed to the solutions of Eq. (\ref{eq2}).
We refer the reader to Ref. \onlinecite{Anc14} and
references therein for technical details on how this equation has been
solved. 

\begin{figure}[t]
\includegraphics[width=0.8\linewidth,clip=true]{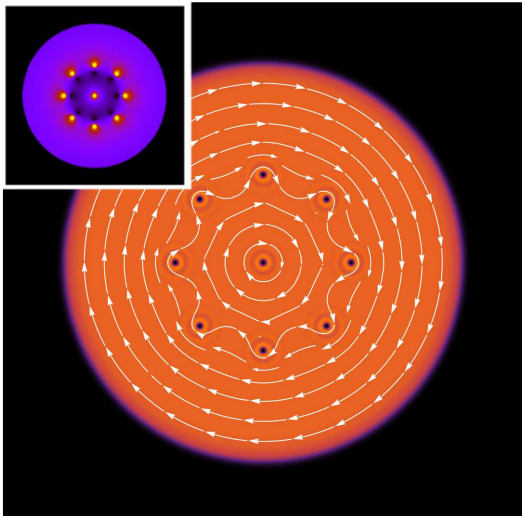}
\caption{\label{fig5} 
Circulation lines  of the velocity field corresponding to the  
$N_v = 9$  configuration of Fig. \ref{fig1}.
The inset displays in a color scale the regions around the vortex cores 
where the modulus of the velocity field is higher
(bright spots). 
The dark spots are regions of low vorticity due to the interference between the
velocity fields of neighboring vortices.
}
\end{figure}

Due to the high computational cost of our calculations, 
we have limited this study to a helium droplet made of
$N_{\rm He}=15000$ helium atoms having a radius 
$R=r_0 N_{\rm He}^{-1/3}$ with $r_0=$ 2.22 \AA{}, i.e., $R = 54.7$ \AA{}.
This droplet is still much smaller than the experimental ones, 
which in turn limits the number of hosted vortices.
However, our findings
can be compared with the experimental results on much 
larger droplets once scaled with a dimensionless characteristic
rotational velocity $\Omega$  defined as\cite{Bro80} 

\begin{equation}
\Omega = \sqrt{\frac{M\,\rho_0 R^3}{8\, \gamma}} \, \omega \; ,
\label{eq4}
\end{equation}
where $\rho_0= 0.0218$ \AA$^{-3}$ is the helium atom density 
and $\gamma= 0.274$ K \AA$^{-2}$ is the surface tension of the liquid.
For the $N_{\rm He}=15000$ droplet, $\Omega =1$ corresponds to 
$\omega = 1.13 \times 10^{10}$ s$^{-1}$. 

Figure \ref{fig1} shows configurations hosting 
$N_v=  2, 4, 7,$ and 9 vortex arrays
obtained with $\Omega= 0.43, 0.54, 0.62, $ and $0.69$, respectively. 
Comparing the top and lateral views, it is apparent that the droplet becomes 
increasingly deformed, oblate-like, as $N_v$ (and thus $\Omega $) increases. Also 
apparent is how the droplet surface locally deforms and the vortex lines bend
forced by the physical requirement that
their open ends hit perpendicularly the surface. The bending is 
smaller for larger $N_v$ and, at variance with the 
classical droplet results,\cite{Cha65,Bro80,Hil08}
the droplet becomes ``wheel-like'' as indeed observed in
the experiments.\cite{Gom14} 

By increasing the angular velocity the number of vortices 
that can be stabilized inside the droplet increases.
Eventually, a maximum number of vortices can be 
hosted, above which the rotating droplet 
will no longer be stable. 
For the $N_{\rm He}=15000$  droplet we have found that the maximum 
$N_v$ value is 9.

As in rotating buckets,\cite{Don91} the higher the angular velocity, 
the more packed the vortex array is
around the rotation axis. This leaves a ``strip'' around 
the equator of the droplet free of vortices that
can be clearly appreciated in the $N_v= 9$ case as shown in Fig. \ref{fig5},
where we display several circulation lines
of the superfluid velocity field. 
The inset shows in a color scale the regions around the vortex cores 
where the modulus of the velocity field is higher
(bright spots). 
As expected, the calculated circulation of the velocity 
field of the superfluid along a path surrounding 
the vortex array equals $N_v$, and equals unity around every single vortex.

\begin{figure}[t]
\includegraphics[width=1\linewidth,clip=true]{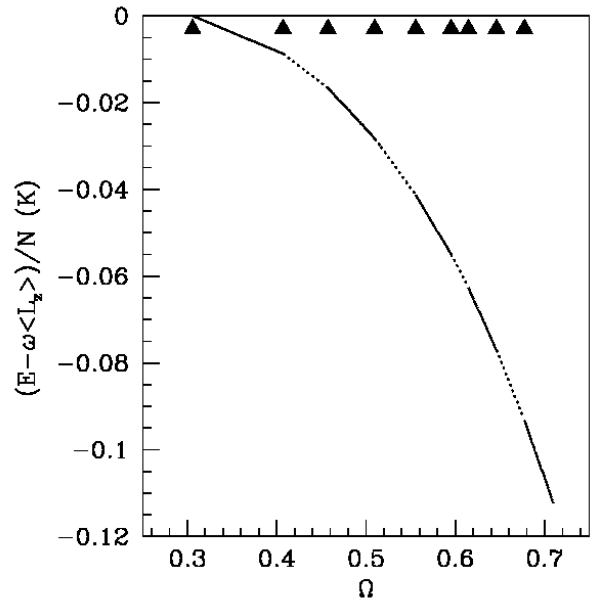}
\caption{\label{fig2}
Stability diagram for a number of vortex lines $N_v = 0,1,2, \ldots, 9$
as a function of the dimensionless angular velocity  $\Omega$.
The zero of the energy scale corresponds to the 
energy of the vortex-free droplet.
The vertical axis is the energy per atom in the corotating frame referred to that
of the vortex-free droplet. 
The triangles mark the crossings between different stability lines.
}
\end{figure}

\begin{figure}[t]
\includegraphics[width=1\linewidth,clip=true]{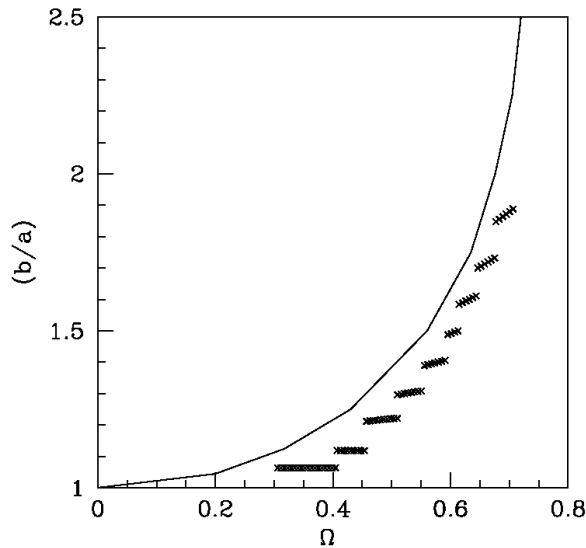}
\caption{\label{fig3} 
Aspect ratio $b/a$ as a function of the dimensionless angular velocity $\Omega$. 
The solid line shows the curve obtained from the
classical model for axisymmetric rotating droplets. 
}
\end{figure}

Figure \ref{fig2} shows the calculated stability diagram. 
As for the rotating bucket,\cite{Anc14,Hes67,Cam79} the 
energetically favored structures for $N_v > 5$ are made of a ring 
of vortices encircling a vortex at the center of the droplet.

\begin{figure}[t]
\includegraphics[width=1\linewidth,clip=true]{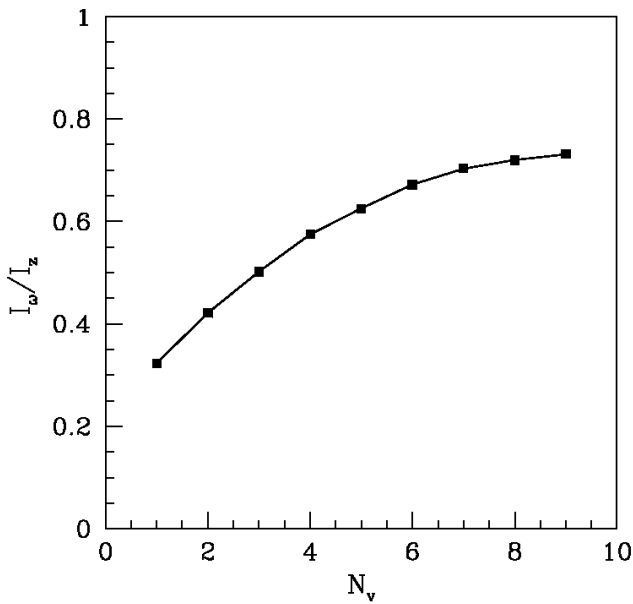}
\caption{\label{fig4} 
Calculated ratio $I_{\omega}/I_z$ shown  
as a function of the number of vortices $N_v$
}
\end{figure}

It is worth observing that  Eq. (\ref{eq1}), which 
strictly applies to an extended vortex triangular (Abrikosov) lattice  
made of a large number of vortex lines, is also fulfilled in the 
present case in spite of the limited number of vortices. 
This occurs in the case of $N_v=7$,
where the equilibrium structure (see Fig. \ref{fig1}), is a ``patch''
of a triangular lattice
whose areal density is $n_v=2/(\sqrt{3}d^2)$,
$d$ being the mean inter-vortex distance.
By equating this expression to Eq. (\ref{eq1})
--with the value $\Omega =0.62$ used to obtain the
7-vortex configuration shown in Fig. \ref{fig1}-- one gets
$d=28.3$ \AA{}.
An average vortex-vortex distance $d=28.2$ \AA{}  can be estimated from Fig. \ref{fig1}, 
which compares very well with the
result of the classical vortex theory.\cite{Cha65,Bro80,Hil08}

Figure \ref{fig1} shows that, disregarding the vortex array, 
the shape of the droplet is almost axially symmetric. 
To determine its $AR$  
we have calculated $a$ and $b$ from the moments of the density 
distribution obtaining\cite{note}
$b/a=[\langle x^2 \rangle/ \langle z^2 \rangle]^{1/2}$,
where $\langle x^2 \rangle =\int{\rho({\mathbf r}) \,x^2} d{\mathbf r}$ and 
$\langle z^2 \rangle =\int{\rho({\mathbf r})\, z^2} d{\mathbf r}$.

The $AR$  
dependence on the angular velocity $\Omega$ is shown in Fig. \ref{fig3},
together with the curve derived from 
the classical model for a rotating liquid droplet\cite{Cha65}, 
and used in Ref.\cite{Gom14} to interpret their data.

The figure shows that for a given angular velocity, the classical 
droplet model overestimates the calculated aspect ratio.
Most likely, the calculated points in Figs. \ref{fig3} 
should get closer to the
classical curve for larger droplets having many vortices,
which unfortunately are beyond the current 
possibilities of the DFT approach.
Notice also that in the experiments of Ref. \onlinecite{Gom14}
axially symmetric stable droplets were observed
with aspect ratios as high as $b/a=2.3$
corresponding to $\Omega=0.71$,  
considerably larger than the shape instability 
threshold of classical droplets leading to 
multi-lobe configurations, $\Omega = 0.56$. 
Our calculations also yield a similar behavior.

It appears from Fig. \ref{fig3} that as $N_v$ increases 
the dependence of the $AR$ on 
$\Omega$ within the corresponding stability region 
(i.e. within each group of crosses shown in Fig.\ref{fig3})
becomes increasingly important, i.e. the droplet
is more easily deformed.
Such increase of the $AR$ proceeds by the flattening 
of the droplet as the vortex cores are pushed, as 
the frequency is increased, towards the center of 
the droplet.

Another interesting difference between classical and superfluid behavior,
which is likely related to the deviations from classical theory just discussed,
emerges if we look at the ratio between the moment 
of inertia around the $z$-axis, $I_z $ calculated from the droplet 
mass distribution, and that obtained from the response of the 
superfluid to rotation, $I_{\omega} = \langle \hat{L}_z \rangle/\omega$. 
The ratio $I_{\omega}/I_z$ is shown in 
Fig. \ref{fig4} as a function of $N_v$ taking for $\Omega$ a 
value in the middle of each stability region. 
One may notice that the higher
the angular velocity the closer the moment
of inertia becomes
to the rigid-body moment of inertia.

To summarize, within DFT we have shown that the shape of rotating 
helium droplets 
hosting a number of vortices
evolves from spheroidal at low angular velocities
to wheel-like at high angular velocities. On the one hand, 
multi-lobe configurations present in 
classical viscid droplets\cite{Hil08} are
hindered by the appearance of vortex arrays
whose regular
distribution is hard to accommodate into a 
peanut-like  (or  higher lobe number) shapes. 
On the other hand, the physical requirement that the 
ends of the vortex lines hit perpendicularly the droplet 
surface favors their parallel alignment for 
large vortex arrays, and hence the
appearance of wheel-like shapes,
as indeed observed in the experiments.
Finally, in spite of the apparent 
differences between normal
and superfluid rotating droplets, the classical relationship 
between the aspect ratio and 
the angular frequency is fairly fulfilled, the classical 
relationship underestimating 
the actual angular frequency by less than 10 \% for the 
relevant, larger vortex arrays. 
Thus, it can be used with some confidence 
in the analysis of the experimental results.

We thank Andrey Vilesov for stimulating discussions.
This work has been performed under Grants No. FIS2011-28617-C02-01
from DGI, Spain (FEDER) and  2014SGR401 from Generalitat de Catalunya.


\begin{thebibliography}{99}

\bibitem{Toe04}
J.P. Toennies and A.F. Vilesov, Angew. Chem. Int. Ed. {\bf 43}, 2622 (2004).

\bibitem{Har95}
M. Hartmann, R.E. Miller, J. P. Toennies, and A.F. Vilesov, 
Phys. Rev. Lett. {\bf 75}, 1566 (1995).

\bibitem{Gre98}
S. Grebenev, J.P. Toennies, and A. Vilesov, Science {\bf 279}, 2083 (1998).

\bibitem{Gom12}
L.F. Gomez, E. Loginov, and A. Vilesov, Phys. Rev. Lett. {\bf 108}, 155302 (2012).

\bibitem{Lat14}
E. Latimer, D. Spence, C. Feng, A. Boatwright, A.M. Ellis, and S. Yang, 
Nano Lett. {\bf 14}, 2902 (2014).

\bibitem{Tha14}
Ph. Thaler, A. Volk, F. Lackner, J. Steurer, D. Knez, W. Grogger, F.
Hofer, and W.E. Ernst, Phys. Rev. B {\bf 90}, 155442 (2014).

\bibitem{Bra13}
N.B. Brauer, S. Smolarek, E. Loginov, D. Mateo, A. Hernando, M. Pi,
M. Barranco, W.J. Buma, and M. Drabbels, Phys. Rev. Lett. {\bf 111}, 153002 (2013).

\bibitem{Lea14}
A. Leal, D. Mateo, A. Hernando, M. Pi, and M. Barranco,
Phys. Chem. Chem. Phys. {\bf 16}, 23206 (2014).

\bibitem{Bau95}
G.H. Bauer, R.J. Donnelly, and W.F. Vinen, J. Low Temp. Phys. {\bf 98}, 47 (1995).

\bibitem{Dal00}
F. Dalfovo, R. Mayol, M. Pi, and M. Barranco, Phys. Rev. Lett {\bf 85}, 1028 (2000).

\bibitem{Leh03}
K.K. Lehmann and R. Schmied, Phys. Rev. B {\bf 68}, 224520 (2003).

\bibitem{Anc03} 
F. Ancilotto, M. Barranco, and M. Pi, Phys. Rev. Lett. {\bf 91}, 105302 (2003).

\bibitem{Sol07}
E. Sola, J. Casulleras, and J. Boronat, Phys. Rev. B {\bf 76}, 052507 (2007).

\bibitem{Vin61}
W.F. Vinen, Proc. Roy. Soc. A {\bf 260}, 218 (1961).

\bibitem{Wil74}
G.A. Williams and R.E. Packard, Phys. Rev. Lett. {\bf 33}, 280 (1974).

\bibitem{Don91}
R.J. Donnelly, {\it Quantized vortices in helium II}, Cambridge Studies
in Low Temperature Physics (Cambridge University Press, Cambridge, U.K. 1991), Vol. 3.

\bibitem{Pit03}
L. Pitaevskii and S. Stringari, {\it Bose-Einstein Condensation},
International Series of Monographs on Physics {\bf 116}
(Clarendon Press, Oxford 2003).

\bibitem{Fet09}
A.L. Fetter, Rev. Mod. Phys. {\bf 81}, 647 (2009).

\bibitem{Gom14}
L.F. Gomez et al,
Science {\bf 345}, 906 (2014).

\bibitem{Cha65}
S. Chandrasekhar, Proc. R. Soc. Phys. Lond. A {\bf 286}, 1 (1965).

\bibitem{Bro80}
R.A. Brown and L.E. Scriven, 
Proc. R. Soc. Phys. Lond. A {\bf 371}, 331 (1980).

\bibitem{Hil08}
R.J.A. Hill and L. Eaves, Phys. Rev. Lett. {\bf 101}, 234501 (2008).

\bibitem{Anc14}
F. Ancilotto, M. Pi, and M. Barranco, Phys. Rev. B {\bf 90}, 174512 (2014).

\bibitem{Fey55}
R.P. Feynman,  Progress in Low Temperature Physics, C.J. Gorter, Editor
(North-Holland Publishing Company, Amsterdam 1955), vol. 1, p. 1.

\bibitem{Boh75}
A. Bohr and B.R. Mottelson,  {\it Nuclear Structure} (W.A. Benjamin Inc. Reading, 
Massachusetts, U.S.A. 1975), Vol. II App. 6A.

\bibitem{Nam96}
S.T. Nam, G.H. Bauer, and R.J. Donnelly, 
J. Korean Phys. Soc. {\bf 29}, 755 (1996).

\bibitem{Anc05}
F. Ancilotto, M. Barranco, F. Caupin, R. Mayol, and M. Pi, 
Phys. Rev. B {\bf 72}, 214522 (2005).

\bibitem{Hes67} 
G.B. Hess, Phys. Rev. {\bf 161}, 189 (1967).

\bibitem{Cam79} 
L.J. Campbell and R.M. Ziff, Phys. Rev. B {\bf 20}, 1886 (1979).

\bibitem{note}
We have verified that this gives the same $AR$ as computed 
by direct inspection of the equidensity contour-line plots
obtained from our calculations.
 
\end{thebibliography}
\end{document}